# Production of Polarized $\tau$ Pairs and Tests of CP Violation Using Polarized $e^{\pm}$ Colliders Near Threshold[*][†]

Yung Su Tsai

*Stanford Linear Accelerator Center*
*Stanford University, Stanford, California 94309*

## ABSTRACT

We consider the production of $\tau$ pairs by electron-positron colliding beams at the maximum cross section near the threshold. At this energy $\tau$ pairs are produced mostly in the $s$-wave which implies that the spin of the $\tau$ pairs are almost always pointing in the beam direction independent of the production angle. When both electrons and positrons are longitudinally polarized in the same direction, for example 90%, one can obtain $\tau$ pairs with 99% polarization in the direction of the polarization vectors of the incident beams. Tests of CP violation and study of the structure of weak interactions using such polarized $\tau$ pairs are discussed.

Submitted to *Physical Review D.*

[*] Work supported by the Department of Energy, contract DE–AC03–76SF00515.
[†] Part of the content of this work was presented at the "Workshop on the Tau-Charm Factory in the Era of B-Factory and CESR at SLAC", August 15–16, 1994, and the "Workshop on Hadron Physics at $e^+e^-$ Colliders at IEHP", Beijing, China, October 14–18, 1994.

hep-ph/9410265

# 1. Introduction

CP violation in the Standard Model is highly *ad hoc* in the sense that it was invented to explain the decay of $k_\ell$ and it arbitrarily assumes that there is no CP violation in the leptonic and the first generation quark vertices. In this paper we propose to test whether there is CP violation in the $\tau$ decay using the proposed Tau-Charm Factory with longitudinally polarized electron and positron beams.

The Tau-Charm Factory [1] is a proposed electron-positron colliding beam machine operating at around 4 GeV in the center of mass where $\tau$ and charm particles have maximum cross sections. When a pair of spin $\frac{1}{2}$ particles are produced near threshold they are produced mostly in the $s$-wave, resulting in polarizations of $\tau^\pm$ both pointing in the same direction [2] along either $e^+$ or $e^-$ depending upon the initial polarization of the incident beams. This is true almost independent of the production angle. We show that at $E = 2.087\ GeV$ for the incident electron in the colliding beam the cross section is maximum, and the $s$-wave production is still dominant. For example, if $e^+$ and $e^-$ are both polarized 90% in the direction of $e^-$ momentum, the $\tau$ pair will be 99% polarized in the direction of $e^-$ momentum.

In Chapter 2 we compute the cross section and the polarization of $\tau^-$ and $\tau^+$ using longitudinally polarized $e^-$ and $e^+$ beams. The cross sections and polarization of $\tau^-$ and $\tau^+$ from the Tau-Charm Factory are compared with those obtainable from the B-Factory.

In Chapter 3, we discuss how these polarized $\tau^\pm$ can be used to test CP violation, CPT violation, and conserved vector current theorem in $\tau^\pm$ decays. We constructed a very generic model of CP violation to investigate many salient features of possible CP violation in the semileptonic decay of $\tau$ into $2\pi$.

In Chapter 4, we generalize the observations made in the previous chapter and devise ways to find CP violation in any CP violating decay mode and any CP violating production mechanism. We also conclude that assuming equal luminosities and initial $e^{\pm}$ polarizations, the Tau-Charm Factory is a factor 7.7 better than the B-Factory for checking CP violation in $\tau$.

## 2. Production of Polarized $\tau^{\pm}$ by Polarized $e^{\pm}$ Colliding Beams

In our problem the mass of the electron can be ignored, the error caused by this approximation can be shown to be $\mathcal{O}(m_e^2/E^2)$ by an explicit calculation, which is $10^{-7}$ in our problem. When the mass is ignored $(1 - \gamma_5)/2$ becomes left (right) handed helicity projection operator for an electron (positron), whereas $(1 + \gamma_5)/2$ becomes right (left) handed helicity projection operator for an electron (positron). $\gamma_5$ commutes with 1, $\gamma_5$ and $\sigma_{\mu\nu}$, but anti-commutes with $\gamma_\mu$ and $\gamma_\mu \gamma_5$, thus in the electron positron annihilation the helicity of $e^+$ and $e^-$ must be opposite to each other in order to annihilate if the current consists of vector and axial vector. The opposite holds for scalar, pseudo scalar or tensor. The standard electroweak interaction has only vector and axial vector interactions if we ignore the contribution from neutral Higgs exchange and $g - 2$ of the electron. The anomalous magnetic moment term is negligible at high energy because its contribution to the cross section is

$$\mathcal{O}\left(\left(\frac{m_e}{2E} \frac{\alpha}{\pi} \ell n \frac{2E}{m_e}\right)^2\right) \sim 10^{-15}$$

of the $\gamma_\mu$ terms. We shall also ignore the possible existence of electric dipole moment of $\tau$ because many people [3] have worked on this problem already. Thus we assume CP conservation in the production of $\tau$ pairs and we deal only with possible CP violation in $\tau$ decay. In Chapter 4 we point out an all-purpose method for detecting CP violation including the one caused by the existence of an electric dipole moment of $\tau$. In this paper we shall also ignore the $Z_0$ exchange diagram that contributes $10^{-3}$ to the polarization. This does not affect the accuracy of our experiment because we cannot measure the polarization of electrons and positrons to this accuracy anyway.

Let $H_1$ and $H_2$ be the helicities of $e^-$ and $e^+$ respectively. Let us write the cross section for $e^+e^- \to \tau^+\tau^-$ as $\sigma(H_1, H_2)$. The argument given above shows that with an accuracy of $10^{-7}$ we have $\sigma(+,+) = 0$ and $\sigma(-,-) = 0$ and only $\sigma(+,-)$ and $\sigma(-,+)$ are not zero.

Suppose there are

$N_{1+}$ electrons with helicity $H_1 = +$,

$N_{1-}$ electrons with helicity $H_1 = -$,

$N_{2+}$ positrons with helicity $H_2 = +$, and

$N_{2-}$ positrons with helicity $H_2 = -1$.

The total number of events is proportional to

$$N_{1+}N_{2-}\sigma(+,-) + N_{1-}N_{2+}\sigma(-,+) . \qquad (2.1)$$

The longitudinal polarizations (not helicities) of electrons and positrons are by definition:

$$w_1 = \frac{N_{1+} - N_{1-}}{N_1} \qquad \text{where } N_1 = N_{1+} + N_{1-} .$$
$$w_2 = -\frac{N_{2+} - N_{2-}}{N_2} \qquad \text{where } N_2 = N_{2+} + N_{2-} .$$

From these four equations we have

$$\frac{N_{1+}}{N_1} = \frac{1+w_1}{2}, \quad \frac{N_{1-}}{N_1} = \frac{1-w_1}{2}, \quad \frac{N_{2+}}{N_2} = \frac{1-w_2}{2}, \quad \frac{N_{2-}}{N_2} = \frac{1+w_2}{2}. \qquad (2.2)$$

Substituting Eq. (2.2) into Eq. (2.1) we obtain

$$\frac{N_1 N_2}{4} \left[ (1 + w_1 w_2) \{\sigma(+,-) + \sigma(-,+)\} + (w_1 + w_2) \{\sigma(+,-) - \sigma(-,+)\} \right] .$$

$$(2.3)$$

From Eq. (2.3), we observe the following:

1. When both electrons and positrons are unpolarized, the cross section is by definition

$$\frac{1}{4} \{\sigma(+,-) + \sigma(-,+)\} . \qquad (2.4)$$

   When only the electron beam is polarized, the cross section is

$$\frac{1}{4} \{\sigma(+,-) + \sigma(-,+)\} + \frac{w_1}{4} \{\sigma(+,-) - \sigma(-,+)\} . \qquad (2.5)$$

   Where both the electron and positron beams are polarized, the cross section is

$$\frac{1+w_1 w_2}{4} \{\sigma(+,-) + \sigma(-,+)\} + \frac{w_1 + w_2}{4} \{\sigma(+,-) - \sigma(-,+)\} . \quad (2.6)$$

2. Comparison of Eqs. (2.4), (2.5) and (2.6) shows that no new physics is obtained by polarizing both beams. However when both beams are polarized and when the polarization of $e^+$ is in the same direction as that of $e^-$, the total number of counts is increased by a factor $(1 + w_1 w_2)$ and the effective polarization is increased from $w_1$ to $(w_1 + w_2)/(1 + w_1 w_2)$. We shall often assume that only the electron is polarized in order to simplify the calculation and discussion. When both $e^{\pm}$ are polarized all we need to do is to multiply the whole expression by a factor $(1+w_1w_2)$ and change $w_1$ to $(w_1+w_2)/(1+w_1w_2)$.

3. The $w_1$ and $w_2$ dependence of the cross section given here is applicable also to $e^+e^- \to Z_0 \to \tau^+ + \tau^-$.

4. If we let $w_1 = w_2 = 0.9$, we obtain $(w_1+w_2)/(1+w_1w_2) = 0.994$.

In this paper we shall not assume the existence of the electric dipole moment of $\tau$, thus $T$ is conserved in the production. When $T$ is not violated, the polarization of $\tau^{\pm}$ cannot have components perpendicular to the production plane, *i.e.* terms proportional to $(\vec{p}_1 \times \vec{p}_-) \cdot \vec{w}$ must be zero, where $\vec{p}_1$ and $\vec{p}_-$ are momenta of $e^-$ and $\tau^-$ respectively, and $\vec{w}$ is the polarization vector of $\tau^-$, because $\vec{p}_1$, $\vec{p}_-$ and $\vec{w}$ all change signs under $T$. There is no complex phase associated with the interaction to allow the existence of such a $T$ violating term. Under CP transformation the polarization of $\tau^-$ turns into polarization of $\tau^+$ denoted by $\vec{w}'$, $w_1 \to w_2$, $\vec{p}_- \to -\vec{p}_+$, and $\vec{p}_1 \to -\vec{p}_2$. Thus

$$\vec{w} = \vec{w}' . \tag{2.7}$$

This statement is true even when $Z_0$ is exchanged.

In this paper we use the convention of my 1971 paper (see Section IV of that paper). We use the three-dimensional vectors $\vec{s}$ and $\vec{w}$ in the rest frame of $\tau^-$ to represent its spin and polarization vectors respectively. $\vec{s}$ is an unit vector whereas $\vec{w}$ is defined as

$$w_i = \frac{\text{Number of } \tau^- \text{ with } \vec{s} = \hat{e}_i - \text{Number of } \tau^- \text{ with } \vec{s} = -\hat{e}_i}{\text{Number of } \tau^- \text{ with } \vec{s} = \hat{e}_i + \text{Number of } \tau^- \text{ with } \vec{s} = -\hat{e}_i} \tag{2.8}$$

$(s_-)_\mu$ is the four vector which becomes $(0, \vec{s})$ in the rest frame of $\tau^-$. We define similar vectors $\vec{s}'$, $(s_+)_\mu$ and $\vec{w}'$ for $\tau^+$. The cross section for producing $\tau^-$ with spin $\vec{s}$ and $\tau^+$ with spin $\vec{s}'$ with initial polarization $w_1$ for $e^-$ and $w_2$ for $e^+$ can be written as:

$$\frac{d\sigma}{d\Omega}(w_1, w_2, \vec{s}, \vec{s}') =$$
$$= \frac{e^4}{(2\pi)^2} \frac{1}{4(p_1 \cdot p_2)} \int \frac{d^3 p_+}{2E} \int \frac{d^3 p_-}{2E} \delta^4(p_1 + p_2 - p_- - p_+)$$

$$\times \frac{1}{4} \text{Tr}(1 + \gamma_5 w_1) \slashed{p}_1 \gamma_\mu (1 + \gamma_5 w_2) \slashed{p}_2 \gamma_\nu$$

$$\times \frac{1}{4} \text{Tr}(1 + \gamma_5 \slashed{s}_-)(\slashed{p}_- + M) \gamma_\nu (1 + \gamma_5 \slashed{s}_+)(\slashed{p}_+ - M) \gamma_\mu$$

$$= \frac{\alpha^2}{16E^2} \beta(1 + w_1 w_2) \left[ \left\{ 1 + \cos^2\theta + \frac{\sin^2\theta}{\gamma^2} \right\} + \left\{ \left(1 - \frac{1}{\gamma^2}\right) \sin^2\theta (s_- \cdot s_+) \right. \right.$$

$$+ \frac{1}{E^2} \left[ 2(p_1 \cdot s_-)(p_1 \cdot s_+) - (p_1 \cdot s_-)(p_- \cdot s_+)(1 + \beta x) \right.$$

$$\left. - (p_1 \cdot s_+)(p_+ \cdot s_-)(1 - \beta x) \right] \Bigg\}$$

$$\left. + \frac{w_1 + w_2}{1 + w_1 w_2} \frac{1}{\gamma E} \left\{ 2(p_1 \cdot s_-) + 2(p_1 \cdot s_-) - (p_- \cdot s_+) - (p_+ \cdot s_-) \right\} \right],$$
$$\tag{2.9}$$

where $x = \cos\theta$, $\gamma = E/M$, and $\beta = (1 - \gamma^{-2})^{0.5}$. We notice that $w_1$, $w_2$, $(p_i \cdot s_-)$ and $(p_i \cdot s_+)$ are pseudoscalars, therefore these quantities have to occur an even number of times in our expression because we are dealing with parity conserving electromagnetic interactions in the production. Parity conservation is violated when $Z_0$ exchange is included. at our energy the correction due to weak interaction is $\mathcal{O}(4E^2/M_z^2) = 10^{-3}$. The first curly bracket in Eq. (2.9) represents the cross section when the final polarizations are not measured, the second curly bracket represents the spin correlation and it was first discussed by the author [2] in 1971 and treated subsequently by many people, so we shall not discuss it here. The third curly bracket contains terms which produce polarization. Since we do not have to observe both polarizations at the same time we let $s_+ = 0$. We can obtain the polarization vector $\vec{w}$ for $\tau^-$ using Eqs. (2.8) and (2.9). For this calculation we shall use the coordinate system shown in Fig. 1. In this frame, for $\vec{s} = \hat{e}_{z'}$, we have

$$s_- = (\beta\gamma, 0, 0, \gamma) \ . \tag{2.10}$$

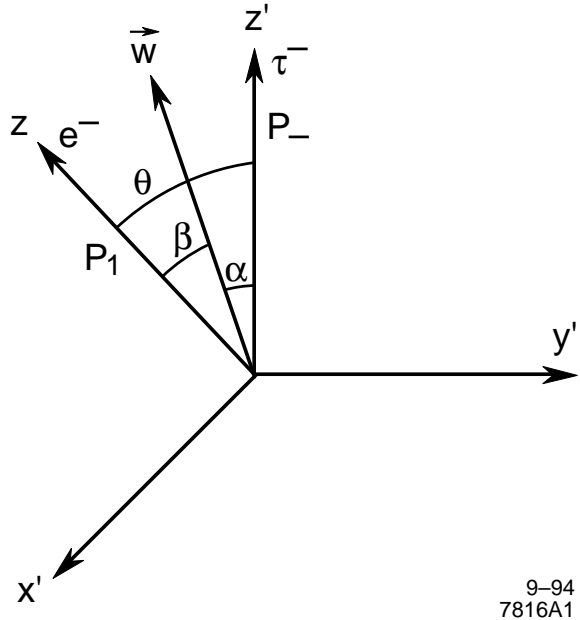

Figure 1. Coordinate system used in calculating the polarization vector $\vec{w}$ for $\tau^-$. $w_y = 0$ because of $T$ invariance in the production of $\tau$ pairs.

For $\vec{s} = \hat{e}_{x'}$, we have

$$s_- = (0, 1, 0, 0) \ . \tag{2.11}$$

$$p_1 = E(1, \sin\theta, 0, \cos\theta) \ . \tag{2.12}$$

$$p_- = E(1, 0, 0, \beta) \ . \tag{2.13}$$

$$p_+ = E(1, 0, 0, -\beta) \ . \tag{2.14}$$

The magnitude of the polarization can be obtained readily from Eqs. (2.8) through (2.14)

$$|\vec{w}| = (w_{x'}^2 + w_{z'}^2)^{1/2} = \left|\frac{w_1 + w_2}{1 + w_1 w_2}\right| \frac{2E\sqrt{p^2 \cos^2\theta + M^2}}{E^2 + M^2 + p^2 \cos^2\theta} \ , \tag{2.15}$$

where $p^2 = E^2 - M^2$. The component of $\vec{w}$ along the $\tau^-$ direction is

$$w_{z'} \equiv |\vec{w}| \cos\alpha = |\vec{w}| \frac{E \cos\theta}{\sqrt{p^2 \cos^2\theta + M^2}} \ . \tag{2.16}$$

The component of $\vec{w}$ along the incident electron direction is

$$w_z \equiv |\vec{w}| \cos\beta = |\vec{w}| \frac{E \cos^2\theta + M \sin^2\theta}{\sqrt{p^2 \cos^2\theta + M^2}} \ . \tag{2.17}$$

Equation (2.15) shows that at $\theta = 0$ or $180°$, the magnitude of the polarization is always maximum independent of energy:

$$|\vec{w}|_{\max} = \left|\frac{w_1 + w_2}{1 + w_1 w_2}\right| \ . \tag{2.18}$$

In Fig. 2a, the magnitudes of the $\tau^\pm$ polarization are plotted assuming $|\vec{w}|_{\max} = 1$ for the Tau-Charm Factory energy $E = 2.087 \ GeV$ and the B-Factory energy $E = 6.0\ GeV$. It is seen that at energy $E = 2.087\ GeV$ where the cross section is maximum, the polarization is almost complete but at the B-Factory energy the polarization is less complete even if the incident electron is completely polarized.

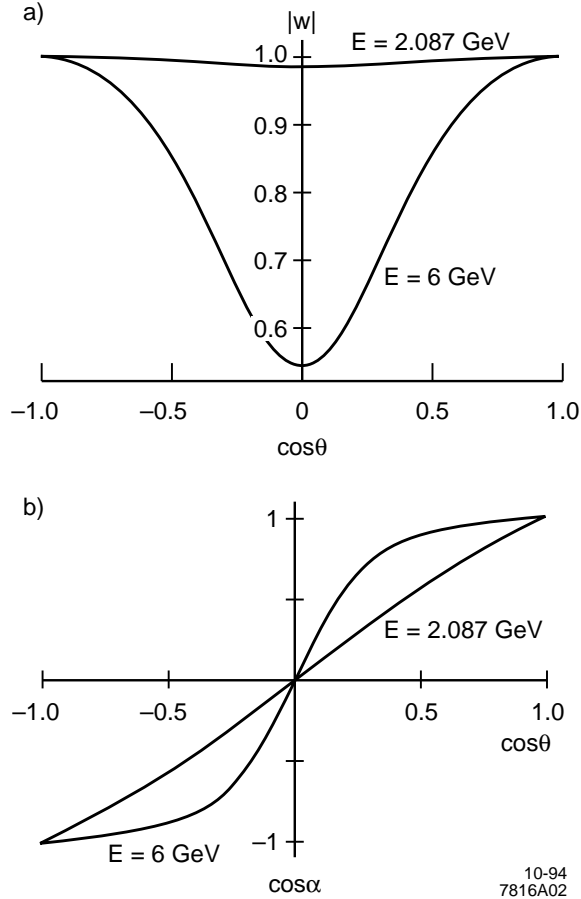

Figure 2. (a) Magnitude of $\tau$ polarization $|\vec{w}|$ as a function of $\cos\theta$ assuming completely polarized electron beam. (b) $\cos\alpha$ versus $\cos\theta$, where $\alpha$ is the angle between $\vec{w}$ (polarization of $\tau$) and $\vec{p}_-$.

In Fig. 2b, the cosine of angle between $\tau^-$ and its direction of polarization is plotted for $E = 2.087$ and $6.0\ GeV$. $\vec{w}$ is almost parallel to the $e^-$ direction if $w_1$ is positive for $E = 2.087\ GeV$ whereas for $E = 6.0\ GeV$ $\vec{w}$ is no longer so parallel to the initial electron polarization because the production is no longer dominated

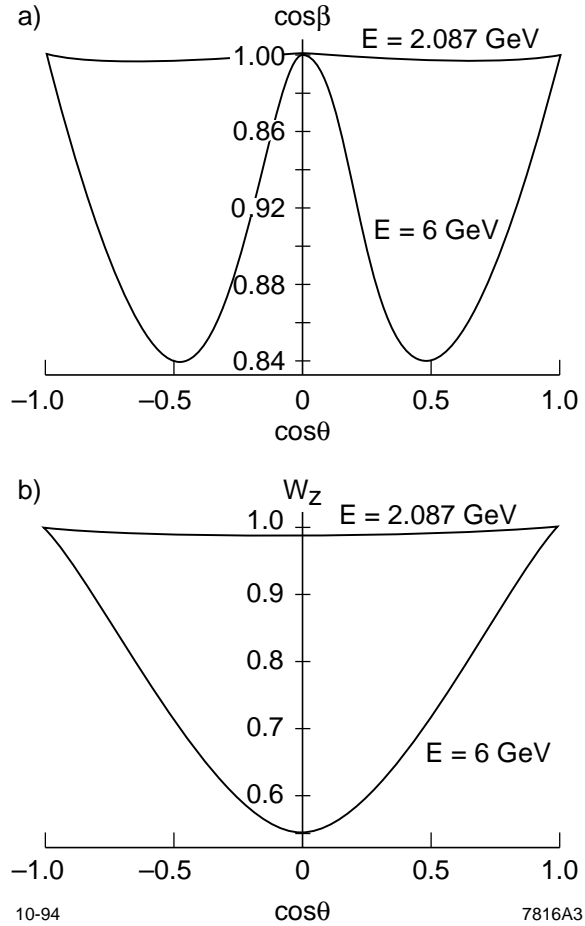

Figure 3. (a) $\cos\beta$ versus $\cos\theta$; (b) $w_z$ is the component of $\tau$ polarization vector along the electron beam direction.

by the $s$ wave.

In Fig. 3a we plot the cosine of the angle between the $\tau^{\pm}$ polarization vector and the incident electron assuming it to have positive helicity. At $\theta = 0°$, $90°$ and $180°$, $\tau^{\pm}$ polarization is always parallel to the electron polarization at all energies. $\cos\beta$ is almost equal to 1 for $E = 2.087\ GeV$ but not quite so for the B-Factory energy.

In Fig. 3b we plot components of $\tau^{\pm}$ polarization along the electron direction assuming the electron to be completely right-handed polarized.

## 2.1 Total Cross Section and Production Rate

The first curly bracket in Eq. (2.9) gives the differential cross section summed over the final spins. Integrating it with respect to solid angle we obtain the total cross section:

$$\sigma(e^+e^- \to \tau^+\tau^-) = \frac{r_e^2 \pi}{6} \left(\frac{m_e}{M}\right)^2 \beta(1-\beta^2)(3-\beta^2)(1+w_1w_2) \ . \tag{2.19}$$

The cross section has a maximum at $\beta = \sqrt{1.5 - \sqrt{1.5}} = 0.5246$ or $E = 2.087 \, GeV$ for $M = 1.777 \, GeV$. When $\beta = 0.5246$ we have $\beta(1-\beta^2)(3-\beta^2) = 1.036$. Let us therefore write $f(\beta) = (1/1.036)\beta(1-\beta^2)(3-\beta^2)$ and $\sigma(e^+e^- \to \tau^+\tau^-) = \sigma_{\max} f(\beta)(1+w_1w_2)$ where

$$\sigma_{\max} = \frac{r_e^2 \pi}{6} \left(\frac{m_e}{M}\right)^2 1.036 = 3.562 \times 10^{-33} cm^2 \ .$$

Table 1

Energy dependence of the cross section for $e^+e^- \to \tau^+\tau^-$.

| $\beta$ | $E(GeV)$ | $f(\beta)$ |
|---|---|---|
| 0.1 | 1.786 | 0.2857 |
| 0.3 | 1.863 | 0.7668 |
| 0.4 | 1.939 | 0.9210 |
| 0.5 | 2.052 | 0.9953 |
| 0.5246 | 2.087 | 1.0000 |
| 0.55 | 2.128 | 0.9988 |
| 0.6 | 2.221 | 0.9785 |
| 0.9951 | 6.0 | 0.1688 |

Table I gives the numerical value of $f(\beta)$. We notice that at the B-Factory energy the cross section is 1/6 that of the maximum cross section at $E = 2.087\ GeV$.

The factor $(1 + w_1 w_2)$ is the spin dependence of the total cross section. When either $w_1 = 0$ or $w_2 = 0$ this factor is one. When $w_1 = w_2 = \pm 1$ this factor is 2. When $w_1 = -w_2 = \pm 1$ this factor is zero. In the circular ring if one waits long enough, positrons (electrons) will be polarized parallel (antiparallel) to the magnetic field, reaching the value 0.924 if the guiding field is uniform. These transverse polarizations can be rotated 90° so that polarizations become longitudinal. In the ideal case we have $w_1 = w_2 = \pm 0.924$. In this case we have $(1 + w_1 w_2) = 1.85$. The time necessary to reach this maximum possible radiative beam polarization is too long with the existing design of the Tau-Charm Factory. The time dependence of the polarization is [4,5] $p(t) = 0.924 \left(1 - e^{-t/T_{pol}}\right)$, where $T_{pol}$ in sec is given by

$$T_{pol}(\sec) = \frac{98.7\, r^2 R}{E^5}$$

where

$E = 2.087\ GeV$, is the beam energy

$r = 12$ meters, is the bending radius

$R = 60$ meters, is the mean radius of the machine.

$T_{pol}$ is approximately 6 hours (which is too long). One can reduce this time by reducing $r$ and $R$ and also by inserting wigglers. Another way to obtain the polarized beam is to inject a polarized electron beam which reaches about 80% polarization at SLAC now but eventually may reach almost [6] 100%. Polarized positrons [7] can be obtained by pair production using high energy circularly polarized photons produced by back scattering of polarized laser beams on high energy electrons.

The design luminosity in 1989 was $10^{33}$ cm$^{-2}$/sec, but now it probably could [8] reach $3 \times 10^{33}$ cm$^{-2}$/sec. Using $10^{33}$ we obtain a rate of 3.56 $(1+w_1 w_2)$ $\tau$ pairs/sec. Thus we obtain $(1 \sim 6) \times 10^8$ $\tau$ pairs/year. This means with several years of running one can obtain a sensitivity of $10^{-4}$ for testing CP violation in the $\tau$ decay. If CP violation in $\tau$ decay is of order $10^{-3}$, similar to the neutral kaon decay, we should be able to investigate the structure of CP violation in $\tau$ decay using the Tau-Charm Factory.

## 3. Tests of CP and CPT Violations in $\tau$ Decay

In quantum mechanics, the time reversal operator, $T$, is the least intuitive among $T$, $C$, and $P$ operators, because under $T$ $i$ must become $-i$ in addition to changing $t$ into $-t$. The requirement of $i$ going into $-i$ can be seen by applying $T$ to the most important commutators in quantum mechanics:

$$[x_i, p_j] = i\delta_{ij} . \qquad (3.1)$$

$T[x_i, p_j]T^{-1} = -[x_i, p_j]$ . Thus the commutation relation, Eq. (3.1), will not be true unless $TiT^{-1} = -i$.

In order to construct a $T$ noninvariant model, we first construct a $T$ invariant interaction with a real coupling constant and then make this real coupling constant complex with a nonvanishing imaginary part.

Let $A = |A|e^{i\delta_w}$ be such a coupling constant with $\delta_w \neq 0$ or $\pi$ for $\tau^-$ decay. We have $TAT^{-1} = A^* \neq A$, thus $T$ is violated in the theory. Testing the existence of $\delta_w$ in the $\tau$ decay is the purpose of this chapter. In quantum mechanics, the overall phase of the matrix element of any process is undetectable because the transition probability is square of the matrix element. Thus the complex coupling constant must be defined with respect to some other coupling constant whose phase is known. Only the interference between the two will produce a $T$ violating effect.

The weak Hamiltonian responsible for $\tau^\pm$ decay can be written in general as

$$H_{\text{weak}} = \sum_i \left( \frac{j_i^+ J_i^-}{q^2 - M_i^2} + \frac{j_i^- J_i^+}{q^2 - M_i^2} \right) , \qquad (3.2)$$

where $j_i^+$ represents a leptonic current whose final charge $-$ initial charge is positive, *i.e.* $\tau^- \to \nu_\tau$, $i$ represents different particles exchanged such as left-handed $W$'s, right-handed $W$'s, charged Higgs, etc. $J_i^-$ is the hadronic or leptonic current whose final charge $-$ initial charge is negative. The first term in Eq. (3.2) gives the decay of $\tau^-$, whereas the second term gives the decay of $\tau^+$. One of the requirements of TCP theorem is that $H_{\text{weak}}$ be Hermitian, and thus the second term is the Hermitian conjugate of the first. Therefore if there is any complex coupling constant in the decay of $\tau^-$, the corresponding coupling constant for the $\tau^+$ decay must be the complex conjugate of the former.

Let $A = |A|e^{i\delta_w}$ be the complex coupling constants responsible for the $T$ noninvariant decay of $\tau^-$, then TCP invariance demands that the coupling constant $\overline{A}$ responsible for the $T$ noninvariant $\tau^+$ decay must be

$$\overline{A} \equiv |\overline{A}|\, e^{i\overline{\delta}_w} = |A|\, e^{-i\delta_w} , \qquad (3.3)$$

which implies $|\overline{A}| = |A|$ and $\overline{\delta}_w = -\delta_w$. If either of these is violated, TCP is violated.

In the semileptonic decay mode of $\tau$ with more than one hadron in the final state, for example $\tau^- \to \nu_\tau \pi^- \pi^0$, we have complex phase due to final state interactions given by Breit-Wigner's formula for the $p$ wave resonance ($\rho$). Because the strong interaction is invariant under charge conjugation this phase is not changed when going from $\tau^- \to \nu_\tau + \pi^- + \pi^0$ to $\tau^+ \to \overline{\nu}_\tau + \pi^+ + \pi^0$. Let the phase shift due to strong interaction be $\delta_s$, we have then for $\tau^-$ decay the phase factor $e^{i(\delta_s + \delta_w)}$, but for $\tau^+$ decay we have $e^{i(\delta_s - \delta_w)}$, if TCP is conserved but $T$ is violated. The existence of the strong phase makes it possible to detect the existence of $\delta_w$ even from seemingly $T$ invariant term such as $\vec{w} \cdot \vec{q}_1$, where $\vec{w}$ is the polarization of $\tau^-$ and $\vec{q}_1$ is the momentum of $\pi^-$.

In the previous chapter we showed that $\tau$ can be polarized almost 100% and its direction of polarization is almost along the beam direction independent of the production angle (see Figs. 2 and 3) at $E = 2.087\ GeV$. We have also shown that the polarization vector for $\tau^-$ and $\tau^+$ are parallel to each other and equal in magnitude as long as CP invariance holds in the production. This extra polarization vector $\vec{w}$ of $\tau^-$ enables us to construct rotationally invariant dot products such as $c_1 \vec{w} \cdot \vec{q}_1$ or $c_2(\vec{w} \times \vec{q}_1) \cdot \vec{q}_2$ where $\vec{q}_1$ and $\vec{q}_2$ are the momenta of decay product of $\tau^-$ and similar quantities $c_1' \vec{w}' \cdot \vec{q}_1'$, $c_2'(\vec{w}' \times \vec{q}_1') \cdot \vec{q}_2'$ where $\vec{w}'$ is the polarization vector of $\tau^+$ and $\vec{q}_1'$ and $\vec{q}_2'$ are the momenta of the charge conjugates of $\vec{q}_1$ and $\vec{q}_2$ respectively.

Under CP we have $\vec{q}_1 \to -\vec{q}_1'$, $\vec{q}_2 \to -\vec{q}_2'$ and $\vec{w} \to \vec{w}'$. Thus $\vec{w} \cdot \vec{q}_1 \to -\vec{w}' \cdot \vec{q}_1'$, $\vec{w} \cdot \vec{q}_2 = -\vec{w}' \cdot \vec{q}_2'$, $(\vec{w} \times \vec{q}_1) \cdot \vec{q}_2 \to (\vec{w}' \times \vec{q}_1') \cdot \vec{q}_2'$, $\vec{p}_1 \to -\vec{p}_2$, and $w_1 \to w_2$ under CP operation. Thus if CP holds we have

$$c_1 = -c_1' \quad \text{and} \quad c_2 = c_2' \tag{3.4}$$

and violation of Eq. (3.4) is violation of CP invariance. $\vec{w} \cdot \vec{q}_1$ is $T$ even and CP odd, thus $c_1 + c_1' \neq 0$ means not only CP violation but also CPT violation for any process which does not have a strong interaction phase such as pure leptonic decay mode and any semileptonic decay with only one hadron, such as $\nu_\tau + \pi$ and $\nu_\tau + k$. In the leptonic decay of $\tau$ there is only one visible final state, thus one cannot construct the triple product $(\vec{w} \times \vec{q}_1) \cdot \vec{q}_2$. Only when the polarization of the final $e$ or $\mu$ is measured, one can test the CP violation from the pure leptonic decay of $\tau$ unless CPT is violated. Similarly any CP violating effect in the decay $\tau \to \nu_\tau + \pi$ or $\tau \to \nu_\tau + k$ means CPT is also violated.

We conclude that only the semileptonic decay modes of $\tau$ with two or more hadronic final particles can exhibit CP violation without violating CPT at the same time. The best candidate is the decay mode $\tau^\pm \to \nu_\tau + \pi^\pm + \pi^0$. Let us investigate this mode in detail and learn several interesting lessons. The lessons learned can obviously be applied to other decay modes.

## 3.1 $\tau^- \to \nu_\tau + \pi^- + \pi^0$ AND $\tau^+ \to \overline{\nu}_\tau + \pi^+ + \pi^0$.

The energy angle distributions of these two decay modes from polarized $\tau$'s had been worked out in detail in my 1971 [2] paper several years before the discovery of the $\tau$. The investigation of possible CP violation using these two decays had been carried out by C. A. Nelson *et al.* [9] using spin correlation methods first proposed in my 1971 paper [2]. Since in the Tau-Charm Factory $\tau^\pm$ can be made highly polarized we do not need to use the spin correlation which requires the detection of twice the number of particles and thus is more complicated. We also note that in our method $s$ and $p$ wave interference in the two $\pi$ state is crucial in untangling the CP violation whereas Nelson *el al.*'s paper does not seem to have any $s$ wave.

The two $\pi$ decay modes have two distinguished advantages. 1. They have the largest branching ratio (25%). 2. It has a two-body (detectable) hadronic final state which has a large phase shift ($\rho$ resonance). This makes it possible to have a coefficient of $\vec{w} \cdot \vec{q}_1$ violating CP invariance without violating TCP invariance. It also enables one to construct a triple product term $(\vec{w} \times \vec{q}_1) \cdot \vec{q}_2$ to test CP invariance. Our investigation is exploratory. We want to know how different types of CP violating terms in various Lagrangians manifest themselves as the CP violating effect in the experiment.

We shall assume that the $\tau$ neutrino mass is either zero or so small that anything that is of order $(m_\nu/m_\tau)^2$ is unobservable experimentally. With this assumption $1-\gamma_5$ and $1+\gamma_5$ are good helicity projection operators for the $\tau$ neutrino states and the matrix element containing $(1-\gamma_5)u(\nu_\tau)$ and that containing $(1+\gamma_5)u(\nu_\tau)$ do not interfere. As mentioned previously the complex coupling constant responsible for CP violation can manifest itself only through interference with other terms which have a real coupling constant. This consideration shows that one cannot obtain a CP nonconserving effect through interference of right-handed current with the left-handed current by assuming that the coupling constant of the former has a weak phase compared with the latter.

The consideration given above also shows that if we limit the weak interaction to be transmitted only by exchange of spin 1 and spin 0 particles, then we have only two possible choices of matrix elements denoted by $M_1$ and $M_2$ (see Fig. 4) that can interfere with the Standard Model matrix denoted by $M_0$:

$$M_0 = \overline{u}(p_2)(\slashed{q}_1 - \slashed{q}_2)(1-\gamma_5)\, u(p_1)\, L \tag{3.5}$$

$$M_1 = \overline{u}(p_2)\left\{P(\slashed{q}_1 - \slashed{q}_2) + S(\slashed{q}_1 + \slashed{q}_2)\right\}(1-\gamma_5)\, u(p_1) \tag{3.6}$$

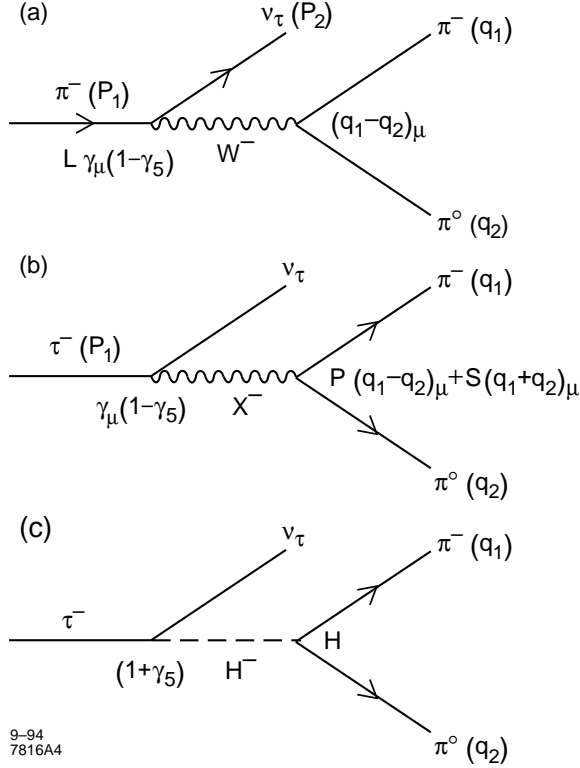

Figure 4. Feynman diagrams for $M_0$, $M_1$, $M_2$ defined in Eqs. (3.5), (3.6) and (3.7); (a) $M_0$: $W^-$ exchange; (b) $M_1$: $X^-$ exchange; (c) $M_2$: $H^-$ exchange.

$$M_2 = \overline{u}(p_2)(1 + \gamma_5)\, u(p_1)\, H \qquad (3.7)$$

$p_1, p_2, q_1, q_2$ are momenta of $\tau^-$, $\nu_\tau$, $\pi^-$ and $\pi^0$ respectively and [2]

$$\begin{aligned} L &= g_{\tau\rho\nu}\, g_{\rho\pi\pi}\, \frac{-1}{(q_1 + q_2)^2 - M_\rho^2 + i\Gamma M_\rho} \\ &= g_{\tau\rho\nu}\, g_{\rho\pi\pi}\, \frac{e^{i\delta_{s1}}}{\sqrt{((q_1 + q_2)^2 - M_\rho^2)^2 + \Gamma^2 M_\rho^2}} \end{aligned} \qquad (3.8)$$

$\delta_{s1}$ is the strong interaction phase shift for the $\pi^-\pi^0$ system in $p$ wave ($\rho$ resonance). Notice that the conserved vector current theorem [2] in the Standard Model says that $\pi^-\pi^0$ cannot be in the $S$ state.

$M_1$ is another left-handed current due to exchange of a higher mass spin 1 particle called $X$. For this current, both $s$ and $p$ waves are allowed for the $\pi^-\pi^0$ system because there is no CVC theorem here and we allow $T$ violating complex coupling constants in $M_1$. The vector particle $X$ couples to all leptons and quarks, probably obeying some yet to be discovered symmetry principle. In our problem $X$ is coupled to both $\tau\nu_\tau$ and the first generation quarks $\bar{u}d$. Thus we will be seeing the combined effect of CP violation in both the $\tau\nu_\tau$ and $\bar{u}d$ sectors. Let the complex weak phase for the $\tau\nu_\tau X$ vertex be $\exp(i\,\delta_{w\tau X})$ and that for the $\bar{u}dX$ vertex be $\exp(i\delta_{w1X})$. Then in our problem only the combination

$$\delta_{wX} \equiv \delta_{w\tau X} + \delta_{w1X} \tag{3.9}$$

will appear.

The term $P$ in Eq. (3.6) contains the same strong interaction phase factor $\exp(i\delta_{s1})$ defined in Eq. (3.8) and thus in the interference between $M_0$ and $M_1$ given by $M_0^+ M_1 + M_1^+ M_0$ this strong interaction phase factor cancels out. Thus the term $P$ does not contribute to the CP violating effect, only the term $S$ in Eq. (3.6) does. The $s$ wave part contains the $I=2$, $J=0$ $\pi^-\pi^0$ phase factor $e^{i\delta_{s0}}$ which is different from the $p$ wave one.

$M_2$ is the matrix element for charged Higgs exchange [10]. The part proportional to $(1-\gamma_5)$ in Eq. (3.7) does not interfere with $M_0$, so we left it out. It has $s$ wave interaction phase factor $\exp(i\delta_{s0})$ and the weak phase factor $\exp(i\delta_{wH})$, where

$$\delta_{wH} \equiv \delta_{w\tau H} + \delta_{w1H} \;, \tag{3.10}$$

where $\delta_{w\tau H}$ is the $T$ violating weak phase associated with $\tau\nu_\tau H$ vertex, while $\delta_{w1H}$ is the similar phase for the first generation quarks. In summary the phases associated with $L$, $P$, $S$ and $H$ are

$$L = |L|\exp(i\delta_{s1}) \tag{3.11}$$

$$P = |P|\exp(i\delta_{s1} + i\delta_{wX}) \tag{3.12}$$

$$S = |S|\exp(i\delta_{s0} + i\delta_{wX}) \tag{3.13}$$

$$H = |H|\exp(i\delta_{s0} + i\delta_{wH}) . \tag{3.14}$$

If CPT invariance holds we have for $\tau^+$ decay

$$\overline{L} = |L|\exp(i\delta_{s1}) \tag{3.15}$$

$$\overline{P} = |P|\exp(i\delta_{s1} - i\delta_{wX}) \tag{3.16}$$

$$\overline{S} = |S|\exp(i\delta_{s0} - i\delta_{wX}) \tag{3.17}$$

$$\overline{H} = |H|\exp(i\delta_{s0} - i\delta_{wH}) . \tag{3.18}$$

Since strong interaction is $C$ invariant, the strong interaction phase shifts $\delta_{s1}$, $\delta_{s0}$ are not changed when going from $\tau^-$ to $\tau^+$ whereas the weak phases $\delta_{wX}$ and $\delta_{wH}$ change sign because of Hermiticity of the Lagrangian, which results in the TCP Theorem. The decay energy-angle distribution of the decay, polarized $\tau^- \to \nu_\tau + \pi^- + \pi^0$ can be written as:

$$\Gamma = \frac{1}{2M_\tau} \frac{1}{(2\pi)^5} \int \frac{d^3p_2}{2E_2} \int \frac{d^3q_1}{2w_1} \int \frac{d^3q_2}{2w_2} \delta^4(p_1 - p_2 - q_1 - q_2) |M_0 + M_1 + M_2|^2 . \tag{3.19}$$

We assume $M_1$ and $M_2$ to be much smaller than $M_0$, therefore we compute: [11]

$$(M_0+M_1+M_2)^+(M_0+M_1+M_2) \cong M_0^+M_0 + (M_0^+M_1 + M_1^+M_0) + (M_0^+M_2 + M_2^+M_0)$$

$$M_0^+ M_0 = 2|L|^2 \frac{\text{Tr}}{4} (1 + \gamma_5 \slashed{w})(\slashed{p}_1 + M)(1 + \gamma_5)(\slashed{q}_1 - \slashed{q}_2)\slashed{p}_2(\slashed{q}_1 - \slashed{q}_2)(1 - \gamma_5)$$

$$= 2|L|^2 \bigg[ 4(w \cdot q_1)M \left\{ (q_1 \cdot q_2) - (p_1 \cdot q_1) + (p_1 \cdot q_2) - m_\pi^2 \right\}$$

$$+ 4(w \cdot q_2)M \left\{ (q_1 \cdot q_2) + (p_1 \cdot q_1) - (p_1 \cdot q_2) - m_\pi^2 \right\}$$

$$+ 4 \bigg\{ -(q_1 \cdot q_2)(p_1 \cdot q_1 + p_1 \cdot q_2 + m_\pi^2) + (p_1 \cdot q_1)^2 + (p_1 \cdot q_2)^2$$

$$- 2(p_1 \cdot q_1)(p_1 \cdot q_2) + m_\pi^2 (p_1 \cdot q_1 + p_1 \cdot q_2) - m_\pi^2 M^2 \bigg\} \bigg] \;.$$

$$(3.20)$$

This gives the energy-angle distribution of $\pi^-$ and $\pi^0$ in the Standard Model which was treated in detail in my 1971 paper [2].

$$M_0^+ M_1 + M_1^+ M_0$$

$$= 2|L|e^{-i\delta_{s1}}\frac{\text{Tr}}{4}(1+\gamma_5\slashed{w})(\slashed{p}_1+M)(1+\gamma_5)(\slashed{q}_1-\slashed{q}_2)\slashed{p}_2$$

$$\times \{P(\slashed{q}_1-\slashed{q}_2)+S(\slashed{q}_1+\slashed{q}_2)\}(1-\gamma_5)$$

$$+ 2|L|e^{i\delta_{s1}}\frac{\text{Tr}}{4}(1+\gamma_5\slashed{w})(\slashed{p}_1+M)(1+\gamma_5)$$

$$\times \{P^*(\slashed{q}_1-\slashed{q}_2)+S^*(\slashed{q}_1-\slashed{q}_2)\}\slashed{p}_2(\slashed{q}_1+\slashed{q}_2)(1-\gamma_5)$$

$$= 4|L|\Bigg[4(w\cdot q_1)M\cos\delta_{wX}(q_1\cdot q_2 - p_1\cdot q_1 + p_1\cdot q_2 - m_\pi^2)|P|$$

$$+ 4(w\cdot q_1)M\cos(\delta_{s0}-\delta_{s1}+\delta_{wX})(q_1\cdot q_2 - p_1\cdot q_1 + m_\pi^2)|S|$$

$$+ 4(w\cdot q_2)M\cos\delta_{wX}(q_1\cdot q_2 + p_1\cdot q_1 - p_1\cdot q_2 - m_\pi^2)|P| \quad (3.21)$$

$$+ 4(w\cdot q_2)M\cos(\delta_{s0}-\delta_{s1}+\delta_{wX})(-q_1\cdot q_2 + p_1\cdot q_2 - m_\pi^2)|S|$$

$$+ (\vec{w}\times\vec{q}_1)\cdot\vec{q}_2 M^2\sin(\delta_{s0}-\delta_{s1}+\delta_{wX})|S|$$

$$+ 4\cos\delta_{wX}\Big\{-(q_1\cdot q_2)(p_1\cdot q_1)-(q_1\cdot q_2)(p_1\cdot q_2)+(q_1\cdot q_2)m_\pi^2$$

$$+ (p_1\cdot q_1)^2 - 2(p_1\cdot q_1)(p_1\cdot q_2)+(p_1\cdot q_1)m_\pi^2 + (p_1\cdot q_2)^2$$

$$+ (p_1\cdot q_2)m_\pi^2 - m_\pi^2 M^2\Big\}|P|$$

$$+ 4\cos(\delta_{s0}-\delta_{s1}+\delta_{wX})\Big\{-(q_1\cdot q_2)(p_1\cdot q_1)+(q_1\cdot q_2)(p_1\cdot q_2)$$

$$+ (p_1\cdot q_1)^2 - (p_1\cdot q_1)m_\pi^2 - (p_1\cdot q_2)^2 + (p_1\cdot q_2)m_\pi^2\Big\}|S|\Bigg]$$

$$M_0^+ M_2 + M_2^+ M_0 = 2|L|\frac{\text{Tr}}{4}(1+\gamma_5\not{w})(\not{p}_1+M)(1+\gamma_5)(\not{q}_1-\not{q}_2)\not{p}_2(1+\gamma_5)H$$

$$+ 2|L|\frac{\text{Tr}}{4}(1+\gamma_5\not{w})(\not{p}_1+M)(1-\gamma_5)\not{p}_2(\not{q}_1-\not{q}_2)(1-\gamma_5)H^*$$

$$= 4|H|\bigg[2(w\cdot q_1)\cos(\delta_{s0}-\delta_{s1}+\delta_{wH})(2(p_1\cdot q_2)-m_\pi^2)$$

$$+ 2(w\cdot q_2)\cos(\delta_{s0}-\delta_{s1}+\delta_{wH})(-2(p_1\cdot q_1)+m_\pi^2)$$

$$+ 4(\vec{w}\times\vec{q}_1)\cdot\vec{q}_2 M\sin(\delta_{s0}-\delta_{s1}+\delta_{wH})$$

$$+ 2M(p_1\cdot q_1 - p_1\cdot q_2)\bigg]$$

(3.22)

## 3.2 Observations

1. The decay energy angle distribution of $\tau^+ \to \bar{\nu}_\tau + \pi^+ + \pi^0$ can be obtained by reversing all momenta of the particles $p_1 \to -p_1'$, $p_2 \to -p_2'$, $q_1 \to -q_1'$, $q_2 \to -q_2'$ and reverse the signs of all weak phases $\delta_{wX} \to -\delta_{wX}$, $\delta_{wH} \to -\delta_{wH}$. When CP is conserved, i.e. $\delta_{wX} = \delta_{wH} = 0$, the coefficients of $w\cdot q_1$ and $w\cdot q_2$ change sign but the coefficients of $(\vec{w}\times\vec{q}_1)\cdot\vec{q}_2$ remain the same under CP operation in agreement with Eq. (3.4). If $\delta_{wX} \neq 0$ or $\delta_{wH} \neq 0$, then Eq. (3.4) is violated thus CP is violated.

2. Only the interference between $s$ wave in $M_{1,2}$ and $p$ wave in $M_0$ contributes to the triple product term $(\vec{w}\times\vec{q}_1)\cdot\vec{q}_2$. Experimentally the existence of this term manifests itself as the asymmetry of $\pi^0$ distribution with respect to the plane formed by $\vec{w}$ and $\pi^-$ momenta. CVC is an exact statement in the Standard Model, thus the existence of the triple product term shows the existence of weak interaction mechanisms other than the Standard Model.

CP is violated if the asymmetry in $\tau^- \to \nu_\tau + \pi^- + \pi^0$ is different from that for $\tau^+$ decay.

3. The $P$ wave part of $M_1$ does not contribute to the observable CP violation because $\cos \delta_{wX} = \cos(-\delta_{wX})$. From this example we can make a very interesting conclusion: Unless two diagrams have two different strong interaction phases, we cannot observe the existence of weak phase using terms involving $w \cdot q_1$ or $w \cdot q_2$. This is because $w \cdot q_1$ and $w \cdot q_2$ are $T$ even in the absence of strong interaction phase differences. Thus we cannot have CP violation without violating CPT using these terms.

4. When the strong interaction phases in $M_0$ and $M_1$ are different the CP violation is proportional to

$$\cos(\delta_{s0} - \delta_{s1} + \delta_{wX}) - \cos(\delta_{s0} - \delta_{s1} - \delta_{wX}) = 2\sin(\delta_{s1} - \delta_{s0})\sin\delta_{wX} \quad (3.23)$$

for the coefficients of $w \cdot q_1$ and $w \cdot q_2$, but

$$\sin(\delta_{s0} - \delta_{s1} + \delta_{wX}) - \sin(\delta_{s0} - \delta_{s1} - \delta_{wX}) = 2\cos(\delta_{s1} - \delta_{s0})\sin\delta_{wX} \quad (3.24)$$

for the coefficients of $(\vec{w} \times \vec{q}_1) \cdot \vec{q}_2$. We notice that when $\delta_{s1} - \delta_{s0} = 0$, Eq. (3.23) is zero whereas Eq. (3.24) is maximum. The physical reason for the former is already explained in point 3 and the reason for the latter is that $(\vec{w} \times \vec{q}_1) \cdot \vec{q}_2$ is $T$ odd. Thus CP violation in this term does not cause violation of CPT even in the absence of strong interactions.

5. Exactly the same observation as point 4 can be made for Eq. (3.22).

6. All observable effects in CP violation can only be produced by the interference between the $p$ wave in $M_0$ and the $s$ wave in $M_1$ and $M_2$ in our model. Our model is generic, so it must be true in general.

## 4. Discussions and Concluding Remarks

Since $\tau^+$ and $\tau^-$ are not observable directly, we have to integrate the production angles and obtain energy-angle distribution of $\pi^-(q_1)$ and $\pi^0(q_2)$ for $\tau^-$ decay and $\pi^+(q_1')$ and $\pi^0(q_2')$ distributions for $\tau^+$ decay. Since we are not doing spin correlation experiments, they do not have to come from the same event. We investigate here features of these energy-angle distributions which will exhibit the CP violation after integrating over $\tau^\pm$ momenta. To simplify the argument let us assume that only the incident electron is polarized. As mentioned in Chapter 2, this does not change any physics. All we need to change is to increase the overall cross section by a factor $(1 + w_1 w_2)$ and replace the electron polarization $w_1$ by $(w_1 + w_2)/(1 + w_1 w_2)$ when positron has a polarization $w_2$.

Let us choose the direction of polarization of $e^-$ as well as its momentum as the $z$ axis and $\pi^-(q_1)$ lies on the $xz$ plane as shown in Fig. 5.

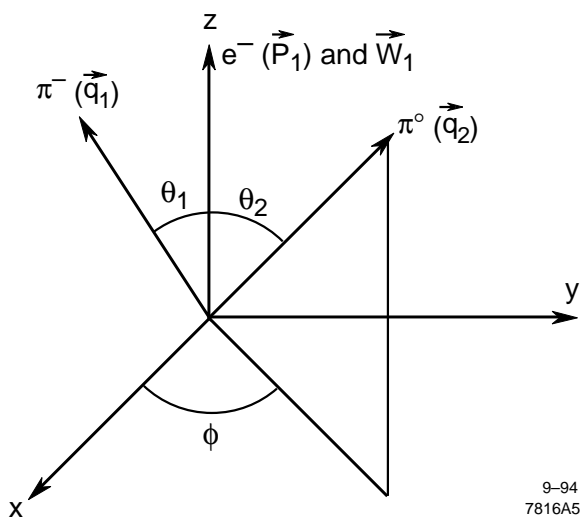

Figure 5. Coordinate system used in Eqs. (4.1) through (4.10).

There are 6 rotationally invariant products involving $\vec{w}_1$:

$$\vec{w}_1 \cdot \vec{q}_1 = w_1 q_{1z} \tag{4.1}$$

$$\vec{w}_1 \cdot \vec{q}_2 = w_1 q_{2z} \tag{4.2}$$

$$\vec{w}_1 \cdot \vec{q}\,'_1 = w_1 q'_{1z} \tag{4.3}$$

$$\vec{w}_1 \cdot \vec{q}\,'_2 = w_1 q'_{2z} \tag{4.4}$$

$$(\vec{w}_1 \times \vec{q}_1) \cdot \vec{q}_2 = w_1 q_{1x} q_{2y} \tag{4.5}$$

$$(\vec{w}_1 \times \vec{q}\,'_1) \cdot \vec{q}\,'_2 = w_1 q'_{1x} q'_{2y} \ . \tag{4.6}$$

Under CP we have

$$\vec{w}_1 \to \vec{w}_2, \quad \vec{q}_1 \to -\vec{q}\,'_1, \quad \vec{q}_2 \to -\vec{q}\,'_2, \quad \vec{p}_1 \leftrightarrow -\vec{p}_2 \ , \tag{4.7}$$

where $\vec{p}_1$ and $\vec{p}_2$ are momenta of electron and positron respectively. We note that $(w_1 + w_2)/(1 + w_1 w_2)$ is symmetric with respect to $w_1 \leftrightarrow w_2$. Let $f_1(q_{1z})$, $f_2(q_{2z})$, $\overline{f}_1(q'_{1z})$, $\overline{f}_2(q'_{2z})$ be the longitudinal distribution of $\pi^-$, $\pi^0$ (from $\tau^-$), $\pi^+$, and $\pi^0$ (from $\tau^+$) respectively. Let $f_3(q_{1x}, q_{2y})$ and $\overline{f}_3(q'_{1x}, q'_{2y})$ be the transverse momentum distributions of $\pi^-\pi^0$ for $\tau^-$ and those of $\pi^+\pi^0$ for $\tau^+$ respectively. If CP is invariant, we have

$$f_1(q_{1z}) = \overline{f}_1(-q'_{1z}) \ , \tag{4.8}$$

$$f_2(q_{2z}) = \overline{f}_2(-q'_{2z}) \ , \quad \text{and} \tag{4.9}$$

$$f_3(q_{1x}, |q_{2y}|) - f_3(q_{1x}, -|q_{2y}|) = \overline{f}_2(q'_{1x}, |q'_{2y}|) - \overline{f}_2(q'_{1x}, -|q'_{2y}|) \ . \tag{4.10}$$

Violation of any one of the equalities in Eqs. (4.8), (4.9) and (4.10) signifies the violation of CP. Nonvanishing of either side of Eq. (4.10) signifies the violation of CVC but does not imply the violation of CP unless the equality is violated. The difference in the detection efficiencies of $\pi^+$ and $\pi^-$ may make Eq. (4.8) rather difficult to verify, but Eq. (4.9) does not have this problem.

As mentioned previously, for leptonic decays or $\tau \to \nu_\tau + \pi$ (or $k$) we cannot have violation of equality like Eq. (4.8) without violating CPT. Thus observation of violation of equality like Eq. (4.8) for these modes is evidence of violation of CPT in these decay modes.

For decays such as $\tau \to \nu_\tau + \pi + k$, $\tau \to \nu_\tau + 3\pi$ we do not have CVC, thus observation of nonvanishing of either side of Eq. (4.10) does not imply violation of the Standard Model. However violation of equality in any one of Eqs. (4.8), (4.9) or (4.10) signifies CP violation in these modes.

Since the derivations of Eqs. (4.8), (4.9) and (4.10) are independent of detail mechanisms of CP violation, they are applicable to all decay channels, as well as all possible CP violations in production of $\tau$'s such as the existence of $\tau$ electric dipole moment. Experimentalists can go ahead and measure the differences between the left and right hand sides of Eqs. (4.8)-(4.10), while theorists can figure out how different models of CP violation will affect the behavior of these functions.

The applications of colliding beams with polarized $e^\pm$ in the production of other particles have not been fully investigated. When hadrons are produced instead of $\tau$'s, their production angles can usually be reconstructed because their decays usually do not involve neutrinos. The method used in Chapter 3 can be used for example in the analysis of $\Lambda\overline{\Lambda}$ and $\xi\overline{\xi}$ production and their decays. The discussions on physics involved in using the transversly polarized $e^\pm$ machine can be found in my 1975 paper [12].

## 4.1 B-Factory versus Tau-Charm Factory for Testing CP in $\tau$ Decay

Let us compare the B-Factory and Tau-Charm Factory for testing CP violation as described in this chapter. Since we are going to integrate with respect to the production angle of $\tau$, we expect the $z$ component of the $\tau$ polarization $w_z$ given by Eq. (2.17) averaged over the differential cross section to give the effective polarization. We obtain from Eqs. (2.17) and (2.9):

$$\overline{w}_z = \int_{-1}^{1} w_z \frac{d\sigma}{dx} dx \bigg/ \sigma = \frac{w_1 + w_2}{1 + w_1 w_2} \frac{1 + 2a}{2 + a^2} \equiv \frac{w_1 + w_2}{1 + w_1 w_2} F(a) \qquad (4.11)$$

where $a = M/E$. $a = 0.8514$ and $0.2961$ respectively for $E = 2.087$ and $6.0\ GeV$; and for $E = 2.087\ GeV$ we have $F(0.8514) = 0.992$, and for $E = 6.0\ GeV$ we have $F(0.2961) = 0.763$. We note $F(1) = 1$ and $F(0) = 0.5$. The total cross section is given by Eq. (2.19) which has a factor $(1 + w_1 w_2)$ that cancels out with the denominator in Eq. (4.11).

Finally the overall merit factor for each machine is

$$\text{Merit} = \text{Luminosity} \times \overline{w}_z \times \text{total cross section}$$
$$\propto \text{Luminosity} \times (w_1 + w_2) \times \sqrt{1 - a^2}\, a^2 (1 + 2a), \qquad (4.12)$$

where $a = M/E$.

Thus if electron and positron are unpolarized, i.e. $w_1 = w_2 = 0$, it has zero value. Assuming the luminosity and the initial beam polarization to be the same for the two machines, the merit factor is determined by the function $f_m(a) = \sqrt{1 - a^2}\, a^2 (1 + 2a)$. For the Tau-Charm Factory we have $f_m(0.8514) = 1.0276$ whereas for the B-Factory we have $f_m(0.2961) = 0.1333$. Thus the Tau-Charm Factory is better than the B-Factory by a factor 7.7 if both have the same luminosity and the initial beam polarizations.

ACKNOWLEDGMENTS

The author wishes to thank Professor T. D. Lee for making a critical remark, and J. D. Bjorken, Richard Prepost, and David Hitlin for useful discussions.